\documentclass[pra, onecolumn, showpacs,aps]{revtex4}
\usepackage[dvips]{graphicx}
\usepackage{bm}

\usepackage{amssymb}
\usepackage{amsmath,amsthm,amssymb}
\usepackage{cases}
\usepackage[mathscr]{euscript}
\usepackage{graphicx}
\usepackage[all]{xypic}\begin{document}
\newtheorem{exercice}{Exercice}[section]
\newtheorem{lemme}{Lemma}[section]

\newtheorem{exemple}{Example}[section]
\newtheorem{théorème}{Theorem}[section]
\newtheorem{definition}{Définition}[section]

\newtheorem{remarque}{Remarque}[section]
\newtheorem{corollaire}{Corollary}[section]
\newtheorem{proposition}{Proposition}[section]

\newenvironment{preuve}{{\bf Proof }}{\hfill $\Box$}
\newenvironment{rcase}
     {\left.\begin{aligned}}
     {\end{aligned}\right\rbrace}

\title{Spinor algebra and null solutions of the wave equation}

\author{Mohammad Wehbe}
\affiliation{D\'epartement de Math\'ematiques, Universit\'e de Bretagne
Occidentale, \\
6 Avenue Le Gorgeu, B.P. 452, 29285 Brest, France.}

\begin{abstract}
In this paper we exploit the ideas and formalisms of twistor theory, to show
how, on Minkowski space, given a null solution of the wave equation, there are precisely two null directions in $\ker df$, at least one of
which is a shear-free ray congruence. \\

\vspace{0.5cm}

\end{abstract}


\maketitle

\section*{Introduction}

\quad Harmonic morphisms are mappings which pull back local harmonic functions
to harmonic functions. On a Riemannian or semi-Riemannian manifold ,they can be
characterized as harmonic maps which enjoy an extra property called horizontal
weak conformality or semi-conformality \cite{Fuglede}, \cite{Fugledee}.\\

If $\varphi: (M,g)\rightarrow (N,h)$ is a mapping between Riemannian manifolds,
then $\varphi$ is semi-conformal if and only if for each $x\in M$, there exists
a function $\lambda:M\rightarrow \mathbb{R}(\neq 0)$ such that $d\varphi\circ
d\varphi^*:T_{\varphi(x)}N\rightarrow T_{\varphi(x)}N$ is $\lambda^2(x)id$,
where $d\varphi^*$ denotes the adijoint of $d\varphi$. In the case of a map
$\varphi:U\subset \mathbb{R}^n\rightarrow\mathbb{C}$, this is equivalent to the
equation
$$\sum\limits_{i=1}^m\left(\dfrac{\partial \varphi}{\partial x_i}\right)^2=0.$$
On Minkwoski space $\mathbb{M}$ endowed with standard coordinates
$(t,x_1,x_2,x_3)$, a harmonic morphism $f:U\subset \mathbb{M}\rightarrow
\mathbb{C}$ corresponds to a null solution of the wave equation:
\begin{eqnarray} (\partial_t f)^2- (\partial_1 f)^2-(\partial_2
f)^2-(\partial_3 f)^2=0\nonumber
\\\partial_{tt}^2f-\partial_{11}^2f-\partial_{22}^2f-\partial_{33}^2f=0.\nonumber
\end{eqnarray}
The first equation can be interpreted as the semi-conformality of $f$ with
respect to the Minkowski metric $g=-dt^2+dx_1^2+dx_2^2+dx_3^2$, and the second
its harmonicity.\\

Our aim in this paper is to give a direct proof that such a mapping determines
a shear-free ray congruence. By the Kerr theorem, thus latter object
corresponds to a complex analytic surface in twistor space \cite{Rindler}.\\
In fact, our proof provides more precise information, specifically, we show
there are exactly two null directions in $\ker df$, at least one of which is
geodesic and shear free. Furthermore, we only require that $f$ be of class
$C^2$. The correspondence between a null solution of the wave equation and a
shear-free ray congruence has already been noted in \cite{BW}, their proof
relies on the analyticity of $f$, there by analytically countinuing $f$ to
Riemannian $\mathbb{R}^4$, one can then apply a theorem of J.C. Wood
\cite{wood} to show that such a mapping determines an integrable complex
structure, which one more corresponds to a complex analytic surface in twistor
space.\\

In more generality, Wood shows that, if $(M^4,g)$ is an oriented Einstein
4-manifold and $\varphi: (M^4,g)\rightarrow (N^2,h)$ a harmonic morphism with a
condition on the set of critical points, then $\varphi$ determines two almost
Hermitien structures on $(M^4,g)$, at least one of which is integrable.
Conversely, if $(M^4,g)$ is also anti-self-dual, then any Hermitien structure
determines a harmonic morphism in a surface.\\

Our result should be seen as the direct analogue of this, with a shear-free ray
congruence on Minkowski space is precisely an integrable complex structure
on a domain of $\mathbb{R}^4$. Our perspective will provide some insight into
the case of more general curved space times, where the methods of twistor
theory no larger apply.

\section{\textbf{Spinorial formulation for null solutions to the wave equation on Minkowski
space}}\label{section1} We consider Minkowski space $\mathbb{M}$ with its
standard metric. Vectors $(x^a)$ may be expressed in terms of spinors by the
correspondence \cite{Rindler}:
\begin{center}
$(t=x^0,x^1,x^2,x^3)\leftrightarrow\dfrac{1}{\sqrt{2}}\left(\begin{array}{cc}t+x^1&x^2+ix^3\\x^2-ix^3&t-x^1\end{array}\right)=(x^{AA'});$
\end{center}
the spinor covariant derivatives $\nabla_{AA'}$ are then given by
\begin{center}
$\nabla_{AA'}$$=$$ \frac{1}{\sqrt2}\left(\begin{array}{cc}
\partial_0+\partial_1&\partial_2-i\partial_3\\\partial_2+i\partial_3&\partial_0-\partial_1\end{array}\right)=\left(\begin{array}{cc}
\nabla_{00'}&\nabla_{01'}\\\nabla_{10'}&\nabla_{11'}\end{array}\right),$
\end{center}
where $\partial_a=\dfrac{\partial}{\partial x^a}$ (Indices are raised and
lowered using the skew forms $\epsilon^{AB}=\epsilon
^{AB}=\begin{pmatrix}0&1\\-1&0\end{pmatrix}$). Note that $(x^a)$ is a real vector if and only if the matrix $(x^{AA'})$ is Hermitian.\\\\
Now let $\varphi:\mathbb{M}\rightarrow \mathbb{C}$ be a $C^2$-mapping, then
$\varphi$ is semi-conformal if and only if
\begin{equation}\label{eq1}
-\left(\frac{\partial\varphi}{\partial
x^0}\right)^2+\sum\limits_{a=1}^3\left(\frac{\partial
\varphi}{\partial x^a}\right)^2=0,
\end{equation}
and $\varphi$ is harmonic, equivalently, $\varphi$ satisfies the wave equation,
if and only if
\begin{equation}\label{eq2}
-\frac{\partial^2\varphi}{(\partial
x^0)^2}+\sum\limits_{a=1}^3\frac{\partial^2\varphi}{(\partial
x^a)^2}=0.
\end{equation}
Equation (\ref{eq1}) is equivalent to the condition $\det\nabla_{AA'}\varphi=0$,
so we deduce that $\varphi$ is semi-conformal if and only if
\begin{equation}\label{eq3}
\nabla_{AA'}\varphi=\xi_A\eta_{A'},
\end{equation}
for some spinor fields $\xi_A$, $\eta_{A'}$ defined on $\mathbb{M}$. One can now show that (\ref{eq1}) and (\ref{eq2}) are equivalent to the pair of
spinor equations \cite{Baird}, \cite{B}:

\begin{center}$\begin{cases}
\nabla_{AA'}\xi^A\eta^{B'}=0&\\\nabla_{AB'}\xi^C\eta^{B'}=0.&
\end{cases}$
\end{center}
Note that these equations imply the integrability of the spinor field
$\xi_A\eta_{A'}$; thus there is locally defined a mapping $\varphi$ such that
$\nabla_{AA'}\varphi=\xi_A\eta_{A'}$. In Minkowski space, a vector $v=(v_0,v_1,v_2,v_3)\in \mathbb{M}$ is null if:
$$-v_0{}^2+v_1{}^2+v_2{}^2+v_3{}^2=0,$$
in which case,
$$v^{AA'}=\lambda\rho^A\bar{\rho}^{A'},$$
for some spinor $\rho^A$ and for some real number $\lambda.$ The following lemma is useful in what follows.
\begin{lemme}\label{lem10}
Let $\xi_A$ non-zero spinor, $(x^a)$ an non-zero vector in
$\mathbb{M}$ and $(x^{AA'})$ its corresponding spinor expression, then:
$\xi_Ax^{AA'}=0$ if and only if there exits a dual spinor $\sigma^{A'}$ such
that $ x^{AA'}=\xi^A\sigma^{A'}.$
\end{lemme}
\noindent \begin{preuve}We know that $\xi_Ax^{AA'}=0$ if and only if
$$\left\{ \begin{array}{cc}\xi_0x^{00'}+\xi_1x^{10'}&=0\\
\xi_0x^{01'}+\xi_1x^{11'}&=0
\end{array}\right.$$
Since $(x^{AA'})\neq 0$, we can assume for example that the component $x^{10'}\neq 0$, hence\\
$$\frac{x^{00'}}{x^{10'}}=\frac{x^{01'}}{x^{11'}}=-\frac{\xi_1}{\xi_0}=\frac{\xi^0}{\xi^1}.$$
It follows that
\begin{center}
$\begin{pmatrix}x^{00'}\\x^{10'} \end{pmatrix}$ =$\sigma^{0'}$
$\begin{pmatrix}\xi^{0}\\\xi^{1} \end{pmatrix}$ and
$\begin{pmatrix}x^{01'}\\x^{11'} \end{pmatrix}$ =$\sigma^{1'}$
$\begin{pmatrix}\xi^{0}\\\xi^{1} \end{pmatrix}$\end{center}
for two numbers $\sigma^{0'}$, $\sigma^{1'}\in\mathbb{C}$, hence
\begin{center}
$x^{AA'}=\xi^A\sigma^{A'}.$
\end{center}
\end{preuve}

Recall that twistor space is the space $\mathbb{T}=\mathbb{C}^4$ equipped with
coordinate $(\xi_A,\eta^{A'})$. The incidence relation \cite{penros}
\begin{equation}\label{eq2.3}
\xi_A=ix_{AA'}\eta^{A'}
\end{equation}
describes the correspondence between points of $\mathbb{T}$ and points of
Minkowski space $\mathbb{M}$. The dual space $\mathbb{T}^*=\mathbb{C}^4$ has coordinates
($\rho^A,\sigma_{A'}$) and given $X^{\alpha}=(\xi_A,\eta^{A'})\in \mathbb{T}$,
we can associate its conjugate
$\bar{X}_{\alpha}=(\overline{\eta}^A,\overline{\xi}_{A'})\in \mathbb{T}^*$,
where
$\bar{\eta}^0=\overline{\eta^{0'}}$, $\bar{\eta}^1=\overline{\eta^{1'}}$,
$\bar{\xi}_{0'}=\overline{\xi_{0}}$, and $\bar{\xi}_{1'}=\overline{\xi_{1}}$
thus
\begin{equation}\label{eq2.4}
\bar{X}_0=\overline{X^2},\qquad \bar{X}_1=\overline{X^3},\qquad
\bar{X}_2=\overline{X^0},\qquad \bar{X}_3=\overline{X^{1}}
\end{equation}
There is a natural product between an element of the twistor space
$X^{\alpha}=(\xi_A,\eta^{A'})\in \mathbb{T}$ and
$L^{\alpha}=(\lambda_A,\nu^{A'})\in \mathbb{T}$ given by:
$$X^{\alpha}\overline{L}_{\alpha}=\xi_A\overline{\nu}^A+\eta^{A'}\overline{\lambda}_{A'}.$$
The correspondence between a light ray in Minkowski space and a point in
twistor space is then given by the following lemma. The proof is standard
linear algebra.
\begin{lemme}\cite{WEHBE}\label{lem11}
Let $X^{\alpha}=(\xi_A,\eta^{A'})$ be a twistor such that $\eta^{A'}\neq 0$,
then $X^{\alpha}$ defines a light ray if and only if
\begin{equation}\label{eq2.5}
X^{\alpha}\overline{X}_{\alpha}=0.
\end{equation}
\end{lemme}

We can give a more invariant picture as follows. Let $\pi$ be the Hopf
fibration defined on $\mathbb{C}\mathbb{P}^3$ with valued in
$\mathbb{H}\mathbb{P}^1\cong S^4$, where $\mathbb{H}$ is the space of the
quaternions $\{z+wj: z,w\in \mathbb{C},ij=-ji, i^2=j^2=-1\}$. Then
$$\pi([f,g,h,k])=[f+gj,h+kj]$$

Let $S^3\subset S^4$ be the equatorial space given by
$S^3=\{(0,x_1,x_2,x_3)\subset S^4\}$, then $\mathcal{N}^5\subset
\mathbb{C}\mathbb{P}^3$ can be identified with $S^3$ by the Hopf fibration:
$\mathcal{N}^5=\pi^{-1}(S^3)$, in particular $[f,g,h,k]\in \mathcal{N}^5$ if
and only if $Re\{\bar{h}f+k\bar{g}+(\bar{h}g-k\bar{f})j\}=0$, i.e. if and only
if
\begin{equation}\label{eq2.1}
\bar{h}f+k\bar{g}+h\bar{f}+\bar{k}g=0.
\end{equation}

To complete the correspondence given by lemma \ref{lem11} by admitting a
twistor of type $X^{\alpha}=(\xi_A,0)$, one comptifies Minkowski space by
adding a light cone at infinity to obtain the space $\mathbb{M}^C$ homeomorphic
to $S^1\times S^3$ \cite{penros}. With this compactification, noting that the
equation (\ref{eq2.5}) is none other than condition (\ref{eq2.1}), we obtain a
correspondence between a point of $\mathcal{N}^5\subset\mathbb{C}\mathbb{P}^3$
and a light ray in $\mathbb{M}^C$, i.e.
$$\mathbb{C}\mathbb{P}^3\supset\mathcal{N}^5=\{\text{light of rays in $\mathbb{M}^C$}\}.$$

Let $\mathcal{S}\subset \mathbb{C}P^3$ a regular complex surface, i.e.
$\mathcal{S}$ is locally parameterized in the form $(z,w)\rightarrow
[\xi_A(z,w),\eta^{A'}(z,w)]$, where $\xi_A(z,w)$, $\eta^{A'}(z,w)$ are
holomorphic in $(z,w)$ and the vectors $(\dfrac{\partial \xi_A}{\partial
z},\dfrac{\partial \eta^{A'}}{\partial z})$, $(\dfrac{\partial \xi_A}{\partial
w},\dfrac{\partial \eta^{A'}}{\partial w})\in \mathbb{C}^4$ are independent for
each  $(z,w)$. We call such a surface a \emph{twistorial surface}.\\\\

Let $\ell$ be a congruence of null curves on a domain $A\subset \mathbb{M}$,
that is $\ell$ is a family of null curves which give a $C^{\infty}$ foliation
of $A$. Let $\xi^A$ be the corresponding spinor field on $A$. Thus at each
point $x\in A$, $\xi^A(x)$ determines the null direction via $\ell$ passing
thought $x$ via the corresponding $v^a\leftrightarrow
v^{AA'}=\xi^A\bar{\xi}^{A'}$. Then the congruence is said to be a
\emph{shear-free ray congruence} (SFR) if and only if each null curve of $\ell$
is geodesic and lie transport of vectors in a 2-dimensional space like
complement of $v$ called the screen space, is conformal, see \cite{Rindler}.
Then $\ell$ is an SFR if and only if
\begin{equation}\label{eq2.8}
\xi^A\xi^B\nabla_{AA'}\xi_B=0.
\end{equation}

We note that this equation depends only on the direction $\xi^0/\xi^1$. Indeed,
writing  $\xi=\xi^0/\xi^1$, the equation (\ref{eq2.8}) is equivalent to:
\begin{center}
$\begin{cases} \xi\nabla_{00'}\xi+\nabla_{10'}\xi=0\\
\xi\nabla_{01'}\xi+\nabla_{11'}\xi=0.
\end{cases}$
\end{center}

\begin{lemme}\label{lem12}
Let $(z,w)\rightarrow [\xi_A(z,w),\eta^{A'}(z,w)]$ be a parameterization of a
twistorial surface $\mathcal{S}$, with $\eta^{1'}(z,w)\neq 0$ for each $z,w$.
Then there exists a local parametrisation of the form
$\widetilde{z}=\widetilde{z}(z,w)$ and $\tilde{w}=\tilde{w}(z,w)$ with respect
which $\mathcal{S}$ is given by
$$(\widetilde{z},\widetilde{w})\rightarrow
[\xi_0(\widetilde{z},\widetilde{w}),\xi_1(\widetilde{z},\widetilde{w}),\widetilde{z},1].$$
\end{lemme}
\noindent \begin{preuve} First, dividing by $\eta^{1'}$, we obtain a
parametrization of $\mathcal{S}$ in the form $$(z,w)\rightarrow
[\xi_0(z,w),\xi_1(z,w),\eta^{0'}(z,w),1].$$
Then we look for a biholomorphic transformation $\tilde{z}=\tilde{z}(z,w)$,
$\tilde{w}=\tilde{w}(z,w)$ such that
$$\frac{\partial \eta^{0'}}{\partial \tilde{w}}=\frac{\partial \eta^{0'}}{\partial z}\frac{\partial z}{\partial \tilde{w}}+
\frac{\partial \eta^{0'}}{\partial w}\frac{\partial w}{\partial
\tilde{w}}=0.$$
By the assumption of regularity
$$rank\left(\begin{array}{ccc}\dfrac{\partial \xi_0}{\partial
z}&\dfrac{\partial \xi_1}{\partial z}&\dfrac{\partial
\eta^{0'}}{\partial z}
\\\\ \dfrac{\partial \xi_0}{\partial w}&\dfrac{\partial \xi_1}{\partial w}&\dfrac{\partial \eta^{0'}}{\partial w}\end{array}\right)=2$$
so that one of the following two conditions is satisfied:
\begin{center} $\left|\begin{array}{cc}\frac{\partial
\eta^{0'}}{\partial z}&\frac{\partial \eta^{0'}}{\partial w}\\
\frac{\partial \xi_0}{\partial z}&\frac{\partial \xi_0}{\partial
w}\end{array}\right|\neq 0,$$ \quad$or$\quad$
$\left|\begin{array}{cc}\frac{\partial \eta^{0'}}{\partial z}&\frac{\partial
\eta^{0'}}{\partial w}\\ \frac{\partial \xi_1}{\partial z}&\frac{\partial
\xi_1}{\partial w}\end{array}\right|\neq 0.$
\end{center}
Suppose that $\left|\begin{array}{cc}\frac{\partial
\eta^{0'}}{\partial z}&\frac{\partial \eta^{0'}}{\partial w}\\
\frac{\partial \xi_0}{\partial z}&\frac{\partial \xi_0}{\partial
w}\end{array}\right|\neq 0$, the other case being similar.\\

Define the map
$\psi:(z,w)\rightarrow(\widetilde{z}=\eta^{0'}(z,w),\widetilde{w}=\xi_0(z,w))$.
Then, since $\left|\begin{array}{cc}\frac{\partial
\eta^{0'}}{\partial z}&\frac{\partial \eta^{0'}}{\partial w}\\
\frac{\partial \xi_0}{\partial z}&\frac{\partial \xi_0}{\partial
w}\end{array}\right|\neq 0$, one can locally find an inverse and so express
$z=z(\widetilde{z},\widetilde{w})$, $w=w(\widetilde{z},\widetilde{w})$. By
construction
$$\frac{\partial \eta^{0'}}{\partial \widetilde{w}}=\frac{\partial \eta^{0'}}{\partial z}\frac{\partial z}{\partial \widetilde{w}}+
\frac{\partial \eta^{0'}}{\partial w}\frac{\partial w}{\partial
\widetilde{w}}=\frac{\partial \widetilde{z}}{\partial
z}\frac{\partial z}{\partial \widetilde{w}}+ \frac{\partial
\widetilde{z}}{\partial w}\frac{\partial w}{\partial
\widetilde{w}}=\frac{\partial \tilde{z}}{\partial
\widetilde{w}}=0.$$

Finally, once more by the regularity, $\eta^{0'}(\widetilde{z})$ is not
constant and can be replaced by $\tilde{z}$ in a neighborhood a point where
$\eta^{0'}(\widetilde{z})\neq0.$

\end{preuve}\\\\

\begin{proposition}\label{th9}
Let $\mathcal{S}\subset\mathbb{C}P^3$ a twistorial surface equipped with a
parameterization $(z,w)\rightarrow [\xi_A(z,w),\eta^{A'}(z,w)]$, and let
\begin{equation}\label{eq2.6}
\xi_A=ix_{AA'}\eta^{A'}
\end{equation}
be the incidence relation. Then, any solution $z=z(x)$ $(x\in \mathbb{M})$ of
the equation (\ref{eq2.6}) is a null solution of the wave equation $\square
z=0$ if and only if $\dfrac{\partial}{\partial
w}\left(\dfrac{\eta^{0'}}{\eta^{1'}}\right)=0$. In particular, any twistor
surface admits such parameterizations and therefore determines an null solution
of the wave equation.
\end{proposition}
\noindent \begin{preuve}
Write $u=x^0+x^1$, $v=x^0-x^1$ and $q=x^2+ix^3$. Then the incidence relation (\ref{eq2.6}) takes the form:\\
\begin{center}
$\left\{ \begin{array}{cc}
r(z,w,x):=u\eta^{0'}+q\eta^{1'}+i\xi_0=0&\\
s(z,w,x):=\bar{q}\eta^{0'}+v\eta^{1'}+i\xi_1=0&
 \end{array}\right.$
 \end{center}
One taking the derivative of these two equations, and writing
$r_w=\dfrac{\partial r}{\partial w}$, $\{r,s\}=r_zs_w-s_zr_w$, we obtain:
\begin{center}
$\dfrac{\partial z}{\partial u}=-\dfrac{s_w\eta^{0'}}{\{r,s\}},\dfrac{\partial
z}{\partial v}=\dfrac{r_w\eta^{1'}}{\{r,s\}}, \dfrac{\partial z}{\partial
q}=-\dfrac{s_w\eta^{1'}}{\{r,s\}},\dfrac{\partial z}{\partial \bar
q}=\dfrac{r_w\eta^{0'}}{\{r,s\}}.$
\end{center}
It follows that
$$\frac{\partial z}{\partial u}\frac{\partial z}{\partial v}-\frac{\partial z}{\partial q}\frac{\partial z}{\partial \bar q}=0.$$
A similar calculation of the second derivatives shows that wave equation is satisfied if and only if
$\{r,s\}(\eta^{1'}\partial_w\eta^{0'}-\eta^{0'}\partial_w\eta^{1'})=0$, from
which the result follows.\end{preuve}

\section{\textbf{Null solutions of the wave equation}}\label{section3}
We study the functions $f:A\subset \mathbb{M}\rightarrow \mathbb{C}$, $A$ open
in $\mathbb{M}$, satisfying the two equations
\begin{equation}\label{eq4}
\begin{cases}
(\partial_t f)^2- (\partial_1 f)^2-(\partial_2 f)^2-(\partial_3
f)^2=0\\\partial_{tt}^2f-\partial_{11}^2f-\partial_{22}^2f-\partial_{33}^2f=0.
\end{cases}
\end{equation}
We call such a function a "null solution" of the wave equation.
\begin{théorème}\label{th1}\cite{BW} Let $f:A\subset \mathbb{M}\rightarrow
\mathbb{C}$ be a $C^2$ null solution of the wave equation (\ref{eq4}). Then there is a pair of spinor field $\xi_A,\eta_{A'}$
such that $\nabla_{AA'}f=\xi_A\eta_{A'}$ which verify
\begin{equation}\label{eq2.10}
\nabla_{AA'}\xi^B\eta^{A'}=\nabla_{AA'}\xi^A\eta^{B'}=0.
\end{equation}

Conversely, any pair of spinor fields $\xi_A,\eta_{A'}$
satisfying (\ref{eq4}), determines a solution of (\ref{eq4}).
\end{théorème}
\begin{théorème}\label{th2}
Let $f:A\subset \mathbb{M}\rightarrow \mathbb{C}$ be a solution of (\ref{eq4}), satisfying $df\neq 0$ at each point of $A$, then the only null fields in $\ker df$
are given by \begin{center} $\lambda\rightarrow\lambda \xi^A\bar{\xi}^{A'}$ et
$\mu\rightarrow \mu \bar{\eta}^A\eta^{A'}$ $(\lambda, \mu \in \mathbb{R}),$
\end{center}and at least one of these fields are tangent to an
SFR. Conversely, let $\xi$ be tangent to an SFR. We define $\xi^0=\xi$,
$\xi^1=1$, $\eta^{0'}=-\nabla_{01'}\xi$, $\eta^{1'}=\nabla_{00'}\xi$. Then
$\xi^A\eta^{A'}$ determines a solution of (\ref{eq4}).
\end{théorème}

 To prove our theorem we require the following lemmas. We note that
for any light ray $v\in\mathbb(M)$, there is a spinor $\rho^A$ with
$v^{(AA')}=\lambda \rho^A \bar{\rho}^{A'} $, where $v^{(AA')}$ is the spinor
expression of $v$.

\begin{lemme}\label{lemme1}
Let $f:A\subset \mathbb{M}\rightarrow \mathbb{C}$ be a $C^2$ solution of (\ref{eq4}). Then there is a pair of spinor fields
$\xi^A,\eta^{A'}$ such that $\nabla_{AA'}f=\xi_A\eta_{A'}$. Moreover, if
$v^{AA'}=\lambda\rho^A\bar{\rho}^{A'}$ is a null direction in $\ker df$ and $df\neq 0$, then, at each point
\begin{itemize}
    \item  either $\rho^A=\alpha \xi^A$,
    \item  or $\rho^A=\beta\overline{\eta^{A'}}$ ($=\beta\bar{\eta}^{A})$,
        with $\alpha$, $\beta$ $\in$ $\mathbb{C}$.

\end{itemize}

\end{lemme}
\noindent \begin{preuve}
The function $f$ is a null solution of the wave equation
(\ref{eq4}), and in particular, a semi-conformal map: thus
$\det(\nabla_{AA'}f)=0$, and we conclude that there exists two spinor field
$\xi_A,\eta_{A'}$ such that
$\nabla_{AA'}f=\xi_A\eta_{A'}$. Suppose that the light ray
$v^{AA'}$ is in $\ker df$, then:
\begin{eqnarray}
\nabla_{AA'}fv^{AA'}=0&\Leftrightarrow& \xi_A\eta_{A'}\rho^A\bar{\rho}^{A'}=0\nonumber\\
&\Leftrightarrow&
\xi_0\eta_{0'}\rho^0\bar{\rho}^{0'}+\xi_0\eta_{1'}\rho^0\bar{\rho}^{1'}+\xi_1\eta_{0'}\rho^1\bar{\rho}^{0'}+\xi_1\eta_{1'}\rho^1\bar{\rho}^{1'}=0\nonumber\\
&\Leftrightarrow&\xi_0\eta_{0'}|\rho^1|^2-\xi_0\eta_{1'}\rho_1\overline{\rho_{0}}-\xi_1\eta_{0'}\rho_0\overline{\rho_{1}}+\xi_1\eta_{1'}|\rho_0|^2=0\nonumber\\
&\Leftrightarrow&\eta_{0'}(\xi_0|\rho_1|^2-\xi_1\rho_0\overline{\rho_1})+\eta_{1'}(\xi_1|\rho_0|^2-\xi_0\rho_1\overline{\rho_0})=0\nonumber\\
&\Leftrightarrow&\frac{\eta_{0'}}{\eta_{1'}}=\frac{\xi_0\rho_1\overline{\rho_0}-\xi_1|\overline{\rho_0}|^2}{\xi_0|\rho_1|^2-\xi_1\rho_0\overline{\rho_1}}=\frac{[\frac{\xi_0}{\xi_1}\rho_1-\rho_0]\overline{\rho_0}}{[\frac{\xi_0}{\xi_1}\rho_1-\rho_0]\overline{\rho_1}}\nonumber\\
&\Leftrightarrow&\frac{\eta_{0'}}{\eta_{1'}}[\frac{\xi_0}{\xi_1}-\frac{\rho_0}{\rho_1}]=\frac{\overline{\rho_0}}{\overline{\rho_1}}[\frac{\xi_0}{\xi_1}-\frac{\rho_0}{\rho_1}].\nonumber
\end{eqnarray}
Then either
\begin{enumerate}
    \item  $$[\frac{\xi_0}{\xi_1}-\frac{\rho_0}{\rho_1}]=0,$$ i.e. $\rho_A=\alpha \xi_A,$
    \item $$\frac{\eta_{0'}}{\eta_{1'}}=\frac{\overline{\rho_0}}{\overline{\rho_1}},$$ i.e. $\rho_A=\beta\overline{\eta_{A'}}$.
\end{enumerate}
\end{preuve}

One writing $v=v^0\partial_0+v^1\partial_1+v^2\partial_2+v^3\partial_3$ for
vector field in $T\mathbb{M}$, we have the corresponding 1-form and its derivative:
\begin{eqnarray}
v_a&=&v^0dx^0+v^1dx^1+v^2dx^2+v^3dx^3.\nonumber\\
dv_a&=&(\partial_1v^0-\partial_0v^1)dx^0\wedge
dx^1+(\partial_2v^0-\partial_0v^2)dx^0\wedge
dx^2+(\partial_3v^0-\partial_0v^3)dx^0\wedge\nonumber\\
&&dx^3+(\partial_2v^1-\partial_1v^2)dx^1\wedge
dx^2+(\partial_3v^1-\partial_1v^3)dx^1\wedge
dx^3+(\partial_3v^2-\partial_2v^3)dx^2\wedge dx^3.\nonumber
\end{eqnarray}
Generally, for a 1-form $\theta=\theta_idx^i$ in a Riemannian manifold $(M,g)$ we have:
$${\rm div}\ \theta=-d^*\theta=g^{ij}\partial_i\theta_j,$$ where $g^{ij}$ is the metric tensor componants.\\
In the Minkowski space $\mathbb{M}$: ${\rm div}\
\theta=\partial_0\theta_0-\partial_1\theta_1-\partial_2\theta_2-\partial_3\theta_3.$
This means that for $v_a=df=\partial_ifdx^i$, we have:
\begin{align}
 {\rm div}\ df&=\partial_{00}^2f-\partial_{11}^1f-\partial_{22}^2f-\partial_{33}^2f\nonumber\\&=\partial_0v^0-\partial_1v^1-\partial_2v^2-\partial_3v^3.\nonumber
\end{align}
By the Poincaré Lemma, if $v_a$ is a 1-form in an open connected set $A\subset\mathbb{M}$, then: $dv_a=0$ if and only if there exist a function $f:A\subset \mathbb{M}\rightarrow \mathbb{C}$ such that $\nabla^{AA'}f=v^{AA'}$. The following Lemma is easily established.
\begin{lemme}\label{lemme3}
The following two conditions are equivalent:\\
a) $dv_a=0$ and ${\rm div}\ v_a=0$\\
b) $\nabla_{AA'}v^{BA'}=0$ and $\nabla_{AA'}v^{AB'}=0$.
\end{lemme}

\emph{Proof of Theorem \ref{th1}}. In fact, if $f$ is a null
solution of the wave equation then from lemma \ref{lemme1}, there exist two
spinor fields $\xi^A$ and $\eta^{A'}$ such that $\nabla_{AA'}f=\xi_A\eta_{A'}$.
The fact that these two fields satisfy (\ref{eq2.10}) is a consequence of lemma
\ref{lemme3}. Conversely, if $\xi_A,\eta_{A'}$ satisfies (\ref{eq2.10}), then
form lamma \ref{lemme3}, there exists a function $f$ such that
$\nabla_{AA'}f=\xi_A\eta_{A'}$
which satisfies (\ref{eq4}).\\

\emph{Proof of Theorem \ref{th2}}. However, the first affirmation
of this Theorem is a consequence of Lemma \ref{lemme1}. We are therefore required to show that at least one of these
null fields is an SFR. However, according to the
Theorem \ref{th1}, we have the two following equations:
\begin{equation}
\nabla_{AA'}\xi^B\eta^{A'}=\nabla_{AA'}\xi^A\eta^{B'}=0\nonumber
\end{equation}
which are equivalent to:
\begin{numcases}{}
\label{eq2.11}\xi^0\nabla_{00'}\eta^{0'}+\eta^{0'}\nabla_{00'}\xi^0+\xi^0\nabla_{01'}\eta^{1'}+\eta^{1'}\nabla_{01'}\xi^0=0\\
\label{eq2.12}\xi^1\nabla_{00'}\eta^{0'}+\eta^{0'}\nabla_{00'}\xi^1+\xi^1\nabla_{01'}\eta^{1'}+\eta^{1'}\nabla_{01'}\xi^1=0\\
\label{eq2.13}\xi^0\nabla_{10'}\eta^{0'}+\eta^{0'}\nabla_{10'}\xi^0+\xi^0\nabla_{11'}\eta^{1'}+\eta^{1'}\nabla_{11'}\xi^0=0\\
\label{eq2.14}\xi^1\nabla_{10'}\eta^{0'}+\eta^{0'}\nabla_{10'}\xi^1+\xi^1\nabla_{11'}\eta^{1'}+\eta^{1'}\nabla_{11'}\xi^1=0\\
\label{eq2.15}\xi^0\nabla_{00'}\eta^{0'}+\eta^{0'}\nabla_{00'}\xi^0+\xi^1\nabla_{10'}\eta^{0'}+\eta^{0'}\nabla_{10'}\xi^1=0\\
\label{eq2.16}\xi^0\nabla_{00'}\eta^{1'}+\eta^{1'}\nabla_{00'}\xi^0+\xi^1\nabla_{10'}\eta^{1'}+\eta^{1'}\nabla_{10'}\xi^1=0\\
\label{eq2.17}\xi^0\nabla_{01'}\eta^{0'}+\eta^{0'}\nabla_{01'}\xi^0+\xi^1\nabla_{11'}\eta^{0'}+\eta^{0'}\nabla_{11'}\xi^1=0\\
\label{eq2.18}\xi^0\nabla_{01'}\eta^{1'}+\eta^{1'}\nabla_{01'}\xi^0+\xi^1\nabla_{11'}\eta^{1'}+\eta^{1'}\nabla_{11'}\xi^1=0.
\end{numcases}
One multiplying (\ref{eq2.11}) by $\xi^1$ and (\ref{eq2.12}) by $\xi^0$ and
subtracting yields:
$$\eta^{0'}\nabla_{00'}\left(\frac{\xi^0}{\xi^1}\right)+\eta^{1'}\nabla_{01'}\left(\frac{\xi^0}{\xi^1}\right)=0.$$
Similary combining the other equations in pairs, one obtains:
\begin{eqnarray}\eta^{0'}\nabla_{10'}\left(\frac{\xi^0}{\xi^1}\right)+\eta^{1'}\nabla_{11'}\left(\frac{\xi^0}{\xi^1}\right)=0\nonumber\\
\xi^0\nabla_{00'}\left(\frac{\eta^{0'}}{\eta^{1'}}\right)+\xi^{1}\nabla_{10'}\left(\frac{\eta^{0'}}{\eta^{1'}}\right)=0\nonumber\\
\xi^{0}\nabla_{01'}\left(\frac{\eta^{0'}}{\eta^{1'}}\right)+\xi^{1}\nabla_{11'}\left(\frac{\eta^{0'}}{\eta^{1'}}\right)=0.\nonumber
\end{eqnarray}
We write $\eta=\eta^{0'}/\eta^{1'}$ and $\xi=\xi^0/\xi^1$, then the above equations are equivalent to:
\begin{numcases}{}
\label{eq2.19}\eta\nabla_{00'}\xi+\nabla_{01'}\xi=0\\
\label{eq2.20}\eta\nabla_{10'}\xi+\nabla_{11'}\xi=0\\
\label{eq2.21}\xi\nabla_{00'}\eta+\nabla_{10'}\eta=0\\
\label{eq2.22}\xi\nabla_{01'}\eta+\nabla_{11'}\eta=0
\end{numcases}
On the other hand, $\xi^A$ is tangent to an SFR if and only if:
\begin{center}
$\begin{cases}
\xi\nabla_{00'}\xi+\nabla_{10'}\xi=0\\
\xi\nabla_{01'}\xi+\nabla_{11'}\xi=0
\end{cases}$
\end{center}
and $\eta^{A'}$ is tangent to an SFR if and only if:
\begin{center}
$\begin{cases}
\eta\nabla_{00'}\eta+\nabla_{01'}\eta=0\\
\eta\nabla_{10'}\eta+\nabla_{11'}\eta=0
\end{cases}$
\end{center}
We take the derivative of  (\ref{eq2.19}) with respect to $\nabla_{11'}$:
\begin{eqnarray}
&&\nabla_{11'}\eta\nabla_{00'}\xi+\eta\nabla_{11'}\nabla_{00'}\xi+\nabla_{01'}\nabla_{11'}\xi=0\quad\text{(by switching the derivatives)}\nonumber\\
&\Rightarrow&-\xi\nabla_{01'}\eta\nabla_{00'}\xi+\eta\nabla_{11'}\nabla_{00'}\xi+\nabla_{01'}(-\eta\nabla_{10'}\xi)=0\quad\text{(by
substituting
 (\ref{eq2.22}) et (\ref{eq2.20}))}\nonumber\\
&\Rightarrow&-\nabla_{01'}\eta(\xi\nabla_{00'}\xi+\nabla_{10'}\xi)+\eta(\nabla_{11'}\nabla_{00'}\xi-\nabla_{01'}\nabla_{10'}\xi)=0\nonumber
\end{eqnarray}
We take the derivative of (\ref{eq2.19}) with respect to $\nabla_{10'}$:
\begin{eqnarray}
&&\nabla_{10'}\eta\nabla_{00'}\xi+\eta\nabla_{00'}\nabla_{10'}\xi+\nabla_{01'}\nabla_{10'}\xi=0\quad\text{(by
switching the
derivatives)}\nonumber\\
&\Rightarrow&-\xi\nabla_{00'}\eta\nabla_{00'}\xi-\eta\nabla_{00'}(\frac{1}{\eta}\nabla_{11'}\xi)+\nabla_{01'}\nabla_{10'}\xi=0\quad\text{(by
substituting (\ref{eq2.21}) et (\ref{eq2.20}))}\nonumber\\
&\Rightarrow&\nabla_{00'}\eta(-\xi\nabla_{00'}\xi+\frac{1}{\eta}\nabla_{11'}\xi)-\nabla_{00'}\nabla_{11'}\xi+\nabla_{01'}\nabla_{10'}\xi=0\nonumber\\
&\Rightarrow&\nabla_{00'}\eta(\xi\nabla_{00'}\xi+\nabla_{10'}\xi)+\nabla_{00'}\nabla_{11'}\xi-\nabla_{01'}\nabla_{10'}\xi=0\quad\text{(par
(\ref{eq2.20}))}.\nonumber
\end{eqnarray}
Thus:
\begin{numcases}{}
\label{eq2.23}-\nabla_{01'}\eta(\xi\nabla_{00'}\xi+\nabla_{10'}\xi)+\eta(\nabla_{11'}\nabla_{00'}\xi-\nabla_{01'}\nabla_{10'}\xi)=0\\
\label{eq2.24}\nabla_{00'}\eta(\xi\nabla_{00'}\xi+\nabla_{10'}\xi)+\nabla_{11'}\nabla_{00'}\xi-\nabla_{01'}\nabla_{10'}\xi=0
\end{numcases}
In particular:
\begin{eqnarray}
&\text{either}\quad&
\left|\begin{array}{cc}-\nabla_{01'}\eta&\eta\\\nabla_{00'}\eta&1\end{array}\right|=-\nabla_{01'}\eta-\eta\nabla_{00'}\eta=0\nonumber\\
&\text{or either}\quad&
\xi\nabla_{00'}\xi+\nabla_{10'}\xi=\nabla_{11'}\nabla_{00'}\xi-\nabla_{01'}\nabla_{10'}\xi=0.\nonumber
\end{eqnarray}
Now we take the derivative of (\ref{eq2.20}) with respect to $\nabla_{00'}$:
\begin{eqnarray}
&&\nabla_{00'}\eta\nabla_{10'}\xi+\eta\nabla_{10'}\nabla_{00'}\xi+\nabla_{11'}\nabla_{00'}\xi=0\quad\text{(by
switching the
derivatives)}\nonumber\\
&\Rightarrow&-\frac{1}{\xi}\nabla_{10'}\eta\nabla_{10'}\xi+\eta\nabla_{10'}(-\frac{1}{\eta}\nabla_{01'}\xi)+\nabla_{11'}\nabla_{00'}\xi=0\quad\text{(by
substituting (\ref{eq2.21}) et (\ref{eq2.19}))}\nonumber\\
&\Rightarrow&\nabla_{10'}\eta(-\frac{1}{\xi}\nabla_{10'}\xi+\frac{1}{\eta}\nabla_{01'}\xi)-\nabla_{10'}\nabla_{01'}\xi+\nabla_{11'}\nabla_{00'}\xi=0\nonumber\\
&\Rightarrow&\frac{\nabla_{10'}\eta}{\eta}(\frac{1}{\xi}\nabla_{11'}\xi+\nabla_{01'}\xi)+\nabla_{11'}\nabla_{00'}\xi-\nabla_{10'}\nabla_{01'}\xi=0
\quad\text{par (\ref{eq2.20})}.\nonumber
\end{eqnarray}
We take the derivative of (\ref{eq2.19}) with respect to $\nabla_{11'}$ (in another way):
\begin{eqnarray}
&&\nabla_{11'}(-\frac{1}{\eta}\nabla_{01'}\xi)+\eta\nabla_{11'}\nabla_{00'}\xi+\nabla_{01'}(-\eta\nabla_{10'}\xi)=0\nonumber\\
&\Rightarrow&-\frac{1}{\eta}\nabla_{11'}\eta(\nabla_{01'}\xi)+\eta(\nabla_{11'}\nabla_{00'}\xi-\nabla_{01'}\nabla_{10}\xi)+\frac{1}{\xi}\nabla_{11'}
\eta\nabla_{10'}\xi=0\nonumber\\
&\Rightarrow&-\frac{1}{\eta}\nabla_{11'}\eta(\nabla_{01'}\xi-\frac{\eta}{\xi}\nabla_{10'}\xi)+\eta(\nabla_{11'}\nabla_{00'}\xi-\nabla_{01'}
\nabla_{10'}\xi)=0\nonumber\\
&\Rightarrow&-\frac{1}{\eta}\nabla_{11'}\eta(\nabla_{01'}\xi-\frac{1}{\xi}\nabla_{11'}\xi)+\eta(\nabla_{11'}\nabla_{00'}\xi-\nabla_{01'}\nabla_{10'}\xi)
=0.\nonumber
\end{eqnarray}
Hence:
\begin{numcases}{}
\label{eq2.25}-\frac{\nabla_{10'}\eta}{\eta}(\frac{1}{\xi}\nabla_{11'}\xi+\nabla_{01'}\xi)+\nabla_{11'}\nabla_{00'}\xi-\nabla_{10'}\nabla_{01'}\xi=0\\
\label{eq2.26}-\frac{1}{\eta}\nabla_{11'}\eta(\frac{1}{\xi}\nabla_{11'}\xi+\nabla_{01'}\xi)+\eta(\nabla_{11'}\nabla_{00'}\xi-\nabla_{01'}\nabla_{10'}\xi)=0
\end{numcases}
and:
\begin{eqnarray}
&\text{either}\quad&
\left|\begin{array}{cc}\frac{\nabla_{10'}\eta}{\eta}&1\\-\frac{1}{\eta}\nabla_{11'}\eta&1\end{array}\right|
=\nabla_{10'}\eta+\frac{1}{\eta}\nabla_{11'}\eta=0\nonumber\\
&\text{or either}\quad&
\xi\nabla_{01'}\xi+\nabla_{11'}\xi=\nabla_{11'}\nabla_{00'}\xi-\nabla_{10'}\nabla_{01'}\xi=0.\nonumber
\end{eqnarray}

Then if $\eta\nabla_{00'}\eta+\nabla_{01'}\eta\neq 0$, we have
$\xi\nabla_{00'}\xi+\nabla_{10'}\xi=0=\nabla_{00'}\nabla_{11'}\xi-\nabla_{01'}\nabla_{10'}\xi=0$
and then (\ref{eq2.25}) et (\ref{eq2.26}) imply that, either
$\xi\nabla_{01'}\xi+\nabla_{11'}\xi=0$, or
$\nabla_{10'}\eta=\nabla_{11'}\eta=0$. But in the latter case, equations
(\ref{eq2.21}) and (\ref{eq2.22}) show that $\nabla_{00'}\eta=\nabla_{01'}\eta=0$,
which contradicts the assumption $\eta\nabla_{00'}\eta+\nabla_{01'}\eta\neq 0$.
Thus $\xi\nabla_{00'}\xi+\nabla_{10'}\xi=\xi\nabla_{01'}\xi+\nabla_{11'}\xi=0$
and $\xi$ is tangent to an SFR.\\
On the other hand, if $\xi\nabla_{00'}\xi+\nabla_{10'}\xi\neq 0$, we obtain in
the same manner that $\eta$ is tangent to an SFR.\\ Conversely, suppose that
$\xi$ is tangent to an SFR, thus:
\begin{center}
$\begin{cases} \xi\nabla_{00'}\xi+\nabla_{10'}\xi=0\\
\xi\nabla_{01'}\xi+\nabla_{11'}\xi=0.
\end{cases}$
\end{center}
We set $\xi^0=\xi$, $\xi^1=1$, $\eta^{0'}=-\nabla_{01'}\xi$,
$\eta^{1'}=\nabla_{00'}\xi$. Then one can easily check that equations (\ref{eq2.11})
to (\ref{eq2.18}) are satisfied, thus $\xi^A\eta^{A'}$ satisfy (\ref{eq2.10})
and therefore there is a function $ f $ satisfying
$$\nabla_{AA'}f=\xi^A\eta^{A'}$$ which is solution of (\ref{eq4}). The proof is completed.

\begin{exemple}\label{ex13}
Suppose that $z=z(x)$ is a solution of the equation
\begin{equation}\label{eq111}\xi(z).x=1\qquad (x\in\mathbb{M})\end{equation} where $\xi=(\xi_1,\xi_2,\xi_3,\xi_4)$ is holomorphic in $z$,
with $\xi_1{}^2-\xi_2{}^2-\xi_3{}^2-\xi_4{}^2=0$, i.e.
$(i\xi_1)^2+\sum\limits_{j=1}^3\xi_j{}^2=0.$\\
In this case, one can easily check that $z$ is a solution of the equation (\ref{eq4}).\\\\
We can parameterize such $\xi$ in the form \cite{bwood}
$$(i\xi_1,\xi_2,\xi_3,\xi_4)=\frac{1}{2h}(1-f^2-g^2,i(1+f^2+g^2),-2f,-2g)$$
where
$f$, $g$ et $h$ are meromorphic functions on $z$.\\
On differentiating (\ref{eq111}) , we obtain $\xi'(z).x.(\partial z/\partial x_j)+\xi_j=0$,
so that $$\frac{\partial z}{\partial x_j}=-\frac{\xi_j}{\xi'.x},$$
which gives:
\begin{align}
\nabla_{AA'}z=&\frac{1}{\sqrt{2}} \begin{pmatrix}
-(\xi_0+\xi_1)/(\xi'.x)&(-\xi_2+i\xi_3)/(\xi'.x)\\
-(\xi_2+i\xi_3)/(\xi'.x)&(-\xi_0+\xi_1)/(\xi'.x)
\end{pmatrix}\nonumber\\&=
\frac{1}{\sqrt{2}h(\xi'.x)}
\begin{pmatrix}
-i(f^2+g^2)&f-ig\\f+ig&i.
\end{pmatrix}\nonumber
\end{align}
Then we find that:
$$\xi_A=\frac{1}{\sqrt{(\sqrt{2}h)\xi'.x}}\begin{pmatrix}f-ig\\i\end{pmatrix},
\quad
\eta_{A'}=\frac{1}{\sqrt{(\sqrt{2}h)\xi'.x}}\begin{pmatrix}-if+g&1\end{pmatrix}$$
We can easily check that $\xi_A$ et $\eta_{A'}$ satisfy the conditions
\begin{center} $\nabla_{AA'}\xi^B\eta^{A'}=0$ and
$\nabla_{AA'}\xi^A\eta^{B'}=0.$
\end{center}
In this case, $\xi^A$ and $\eta^{A'}$ are both an SFR.
\end{exemple}

\begin{remarque}To extend the proof to varieties more general manifolds than
$\mathbb{M}$, when the derivatives commute, we must introduce curvature terms.
In general then, we can not hoped that the theorem remains true, however, by
analogy with the Riemannian case, we conjecture that the result is true for
Einstein manifold.
\end{remarque}

\end{document}